\documentclass[aps,twocolumn,prb,tightenlines,floatfix,showpacs]{revtex4}
\usepackage{graphicx}
\usepackage[english]{babel}
\usepackage{amsmath}
\usepackage{amssymb}
\usepackage{times}
\usepackage{verbatim}

\begin{document}

\title{Contrasting Nodal and Anti-Nodal Behavior in the Cuprates
Via Multiple Gap Spectroscopies}
\author{Dan Wulin$^{1}$, Chih-Chun Chien$^{1}$,
Dirk K. Morr$^{1,2}$ and K. Levin$^{1}$} \affiliation{$^1$James
Frank Institute and Department of Physics, University of Chicago,
Chicago, Illinois 60637 \\ $^2$ Department of Physics, University of
Illinois at Chicago, Chicago, Illinois 60607}

\date{\today}
\begin{abstract}
Using a precursor superconductivity scenario for the cuprates
we present a theory for the temperature dependent
behavior of the spectral gaps
associated with four distinct spectroscopies:
angle resolved photoemission (ARPES), differential
conductance $dI/dV$, quasi-particle interference spectroscopy,
and the autocorrelated ARPES pattern.
We find good agreement for a range of existing experiments
and make predictions for others. Our theory, which incorporates
the necessary (observed) contrast
between the nodal and anti-nodal
response, shows how different nodal gap shapes are associated with
these alternative spectroscopies.
\end{abstract}
\maketitle

A central focus in current studies of high temperature
superconductivity is on establishing the origin of the pseudogap.
The various theoretical scenarios that have been put forth can be broadly categorized into two schools of thought that
either relate the pseudogap to superconductivity through precursor
pairing or ascribe its origin to a new and competing form of
order\cite{Sawatzky}. Important constraints for these scenarios have been
provided by recent scanning tunneling microscopy (STM) \cite{Seamus,
DavisScience09,PuPa} and angle-resolved photoemission spectroscopy
(ARPES) \cite{KoKh,ShenNature} experiments, which have reported 
pronounced differences in the behavior of the spectral gap along the
nodal and anti-nodal directions, and a deviation of the
superconducting gap below $T_c$ from the symmetry of a monotonic
$d$-wave function (with the exception of the moderately underdoped
cuprates at very low temperatures). Whether these observations are
sufficient to rule out one of the above theoretical proposals for
the origin of the pseudogap, as has been claimed \cite{KoKh,ShenNature},
is currently unclear. The situation has been further complicated by
an ongoing controversy about whether
\cite{ViNowa,JDOSPSEUDOGAP,DavisNature2003} or not \cite{HanKoh} these
various experimental probes yield the same spectral gap.

In this article, we address these issues by studying four different
spectroscopic techniques using a single theoretical framework: the spectral function and the
autocorrelated joint density of states (JDOS) derived from ARPES
experiments, as well as the differential conductance, dI/dV,
and quasi-particle interference spectroscopy, obtained from STM
experiments. For each
technique, we follow the experimental protocols to deduce the
angular dependence of the associated spectral gap. Our theoretical
framework is motivated by the idea that the anomalously short
coherence length of the cuprate superconductors and the concomitant
stronger-than-BCS attractive interactions can lead to preformed
pairs above $T_c$ as the origin of the pseudogap\cite{CSTL05}.
The results of our study are two-fold. \textit{Firstly, we demonstrate that
the robust and salient features of STM and ARPES experiments above and below
$T_c$ can be consistently explained within a preformed pair
scenario}. In particular, we show that the superconducting gap
deviates from the pure $d_{x^2-y^2}$-form, i.e., $(\cos k_x - \cos
k_y)$, 
in the moderately underdoped cuprates.
This deviation
only occurs as the pseudogap state is approached near but below $T_c$, in
agreement with experiment \cite{TaLe,PuPa}. 
In the heavily underdoped cuprates, we show how this
deviation occurs for all temperatures \cite{KoTa} below
$T_c$.  
Moreover, we
find that these ubiquitous deviations in gap shape are tied to the
Fermi arcs\cite{KaCh} observed above $T_c$, which is consistent with recent
experimental STM results \cite{PuPa}. 
\textit{Secondly, we demonstrate
that the superconducting gaps extracted from these different
spectroscopies are only in good quantitative agreement in those
parts of the phase diagram where the spectroscopic gap follows a
$(\cos k_x - \cos k_y)$ form}. However, when the gap deviates from
this form, the four spectroscopies can yield widely different
results, particularly in the nodal region. These findings provide
important insight into the question of consistency between different
spectroscopic probes in the cuprate superconductors.

In the theoretical BCS-BEC crossover scenario considered here, the
pseudogap arises from pre-formed pairs in the normal
state, which become non-condensed pair excitations of the condensate
below $T_c$. This scenario is consistent with the emerging physical
picture \cite{PuPa, KoKh,ShenNature} that pseudogap effects persist
below $T_c$, which differentiates our model \cite{CSTL05,ourarpes}
from other approaches
considering precursor pairing\cite{Sawatzky}. For the moderately underdoped systems, there are indications \cite{ShenNature}
that the ground state gap has conventional $d_{x^2-y^2}$ symmetry, which we presume here. 
As previously discussed \cite{CSTL05,ourarpes}, the condensed and
non-condensed (or preformed) pairs yield two distinct contributions
to the fermionic self energy, given by $\Sigma(K) = \Sigma_{sc}(K) +
\Sigma_{pg}(K)$ with
\begin{equation}
\Sigma_{sc}(K) =\frac{\Delta_{sc}^2 (\bf k)}{\omega+\epsilon_k-\mu}; \quad  \Sigma_{pg}(K)= \frac{\Delta_{pg}^2 (\bf
k)}{\omega+\epsilon_k-\mu+i\gamma}
. \label{eq:1b}
\end{equation}
Here $K$ is a four-vector, and we define $\Delta^2({\bf k},T) \equiv
\Delta_{sc}^2({\bf k}, T) + \Delta_{pg}^2({\bf k},T)$. The momentum
dependence for both $\Delta_{sc}$ and $\Delta_{pg}$ is given by the
same pure $d_{x^2-y^2}$-wave form. The condensed pairs have the
usual BCS self energy contribution, $\Sigma_{sc}$, while the self
energy of the non-condensed pairs $\Sigma_{pg}$ is similar except
for an overall broadening factor $\gamma$. This form for
$\Sigma_{pg}$ has been widely accepted in the cuprate community, in
particular, for the normal state \cite{Normanphenom}.
\begin{figure}
\includegraphics[height=4in,clip] {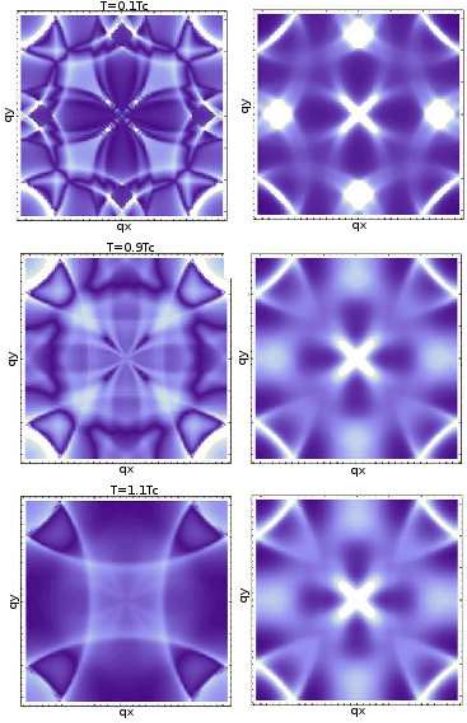}
\caption{(Color online) Each column shows $\delta
n(\textbf{q},\omega)$($J(\textbf{q},\omega)$) for $\omega=-10meV$ and
temperatures $T=0.1,0.9,$ and $1.1T_c$. At low temperatures,
$\delta n$ shows well-defined octet peaks. With increasing temperature
 the octet peaks are broadened
and vanish into the background for $T\geq T_c$. The wavevector $\textbf{q}$ is in units of the inverse lattice constant. }
\label{fig:bigpanel}
\end{figure}

The above framework provides a simple explanation for the different
temperature dependencies of the nodal and anti-nodal spectral gap
observed experimentally.  Above $T_c$, the finite lifetime associated
with the preformed pairs leads to a blurring of the $d$-wave gap
near the nodes and thus to the formation of Fermi arcs
\cite{Chubukov2}. At the anti-nodes, this effect of $\gamma$ is
inconsequential since the normal state gap, as described by
$\Delta_{pg}$, is large there. When the system passes below $T_c$
the nodal region should be thought of as analogous to an ordinary
BCS superconductor which has a gapless normal state and is, thus,
profoundly sensitive to the onset of coherent order (via
$\Delta_{sc}$). The anti-nodal region, with its well developed
normal state gap is, by contrast, rather insensitive to $T_c$.

We next turn to a discussion of the four different spectroscopic
techniques mentioned above. In ARPES experiments, one measures the
electronic spectral function
$A(\textbf{k},\omega)=-\frac{1}{\pi}\mbox{Im}G(\textbf{k},\omega)$,
where $G(\textbf{k},\omega)$ contains the self-energy corrections of
Eq.(\ref{eq:1b}). Derived from the spectral function is the joint
density of states (JDOS), which is given by
\begin{equation}
\label{JDOS} J(\textbf{q},\omega)\propto\displaystyle{\int}d^2k
A(\textbf{k}+\textbf{q},\omega)A(\textbf{k},\omega)
\end{equation}
On the other hand, in STM experiments, one measures the differential
conductance $dI/dV$, which is given by
\begin{equation}
\frac{dI}{dV}(V)\propto-\displaystyle{\int}d \omega f^{\prime}(\omega-V)\displaystyle{\int}\frac{d^2k}{(2\pi)^2}
A (\textbf{k},
\omega)
\end{equation}
Here $f^{\prime}(\omega)$ is the derivative of the Fermi distribution function. Finally, the STM-based quasi particle interference (QPI)
spectroscopy investigates changes in the local density of states
(LDOS) arising from impurity scattering \cite{HanKoh}. For a single point-like
impurity, the Fourier transform of the first order correction to the
LDOS (i.e, the QPI intensity) is given by
\begin{eqnarray}\label{eq:fdos}
\delta n({\bf q},\omega)&\propto&-{\rm Im} \left[ \int
\frac{d^2k}{(2\pi)^2} \left( G({\bf k},\omega)G({\bf k}+{\bf
q},\omega) \right. \right. \nonumber
\\
&& \hspace{-1.5cm} \left. \left. -F_{sc}({\bf k},\omega)F_{sc}({\bf
k}+{\bf q},\omega) -F_{pg}({\bf k},\omega)F_{pg}({\bf k}+{\bf
q},\omega) \right)
\right] \nonumber \\
\end{eqnarray} where
$F_{sc}({\bf k},\omega)\equiv
-\Delta_{sc,\bf k}G({\bf k},\omega)/(\omega+\xi_{\bf k})$
and
$F_{pg}({\bf k},\omega)\equiv -\Delta_{pg,\bf k}G({\bf k},\omega)
/(\omega+\xi_{\bf k}+i\gamma).$
%
\begin{figure*}
\includegraphics[width=2.1in,clip]
{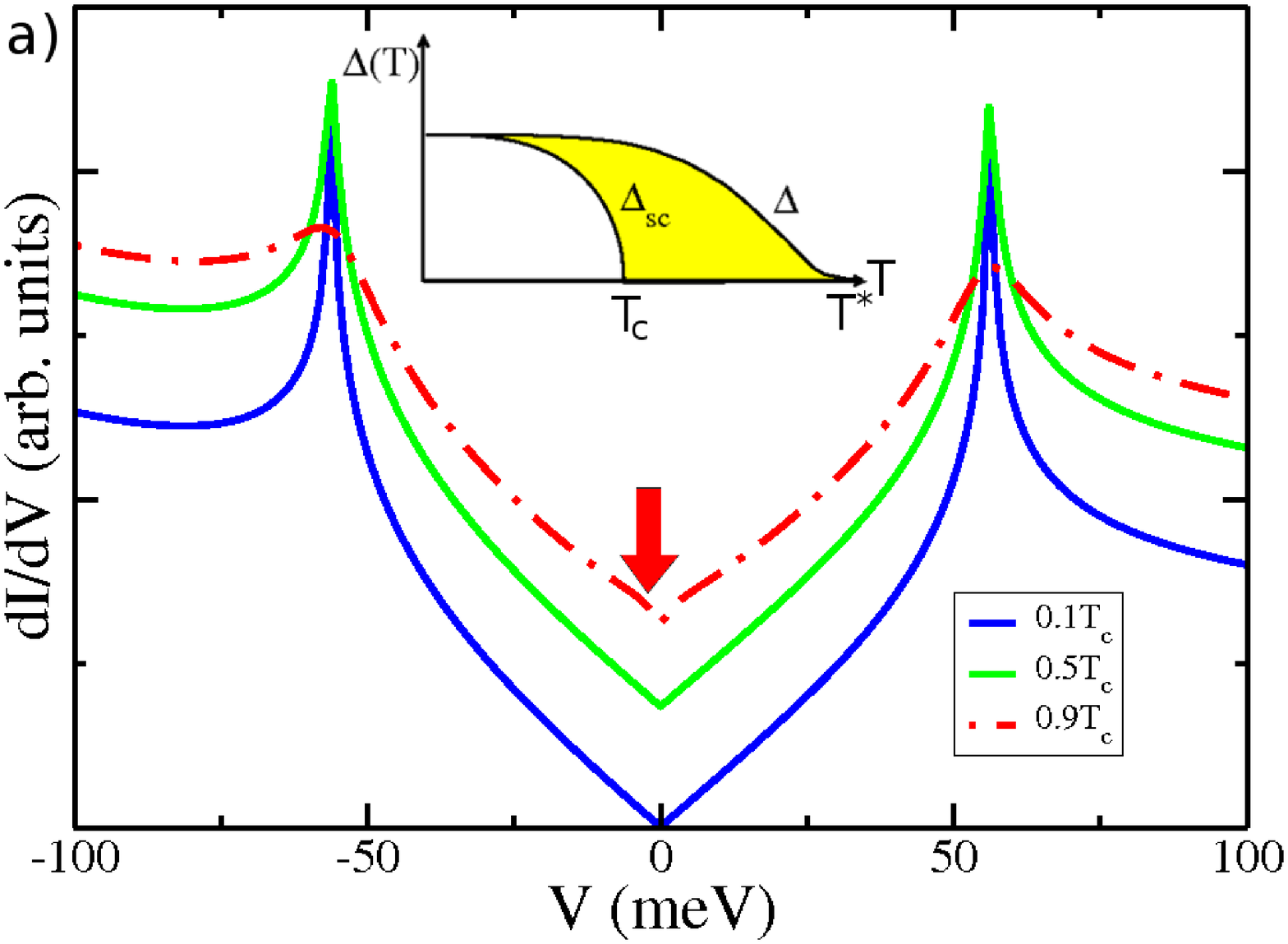}
\includegraphics[width=2.1in,clip]
{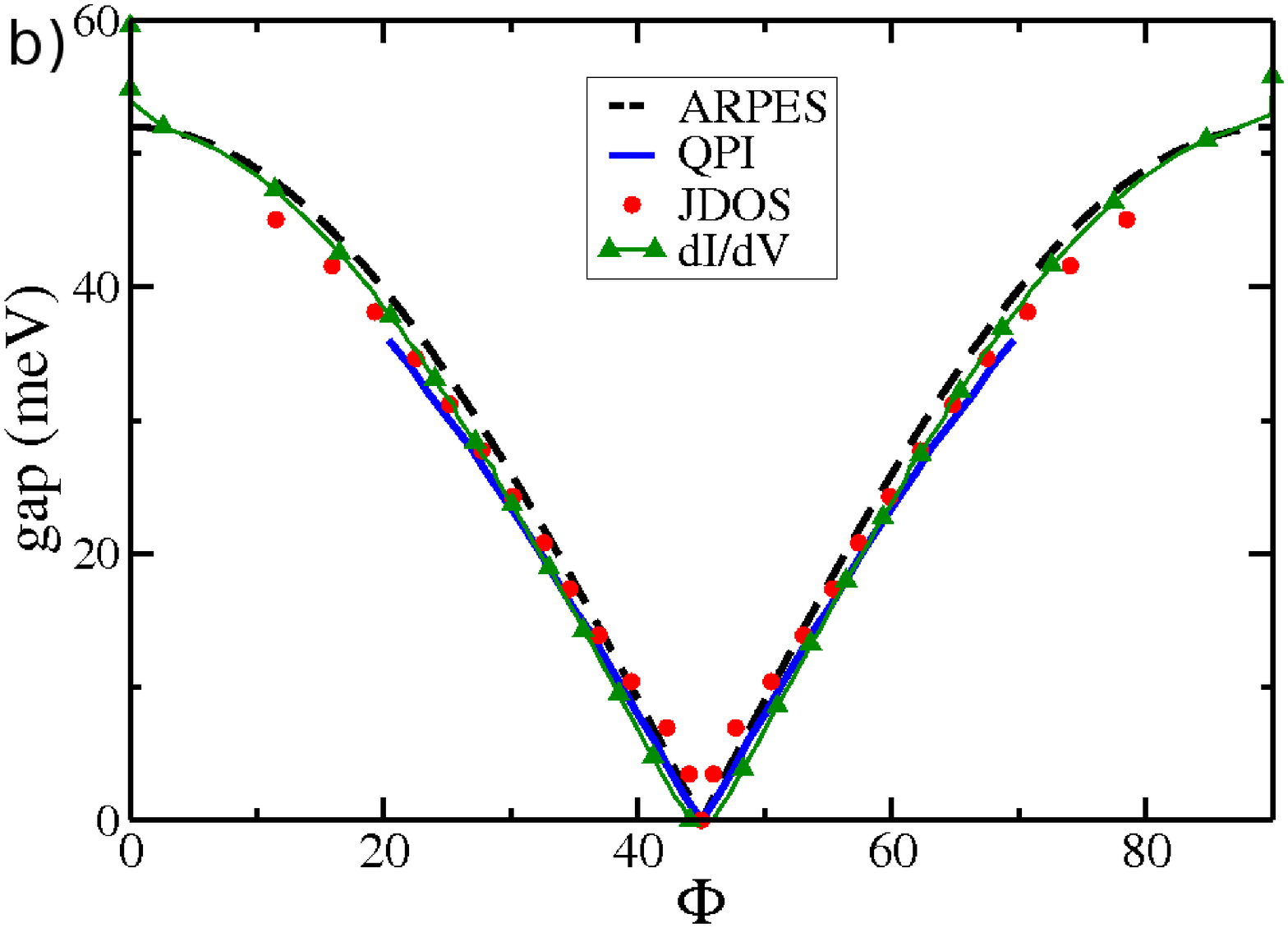}
\includegraphics[width=2.1in,clip]
{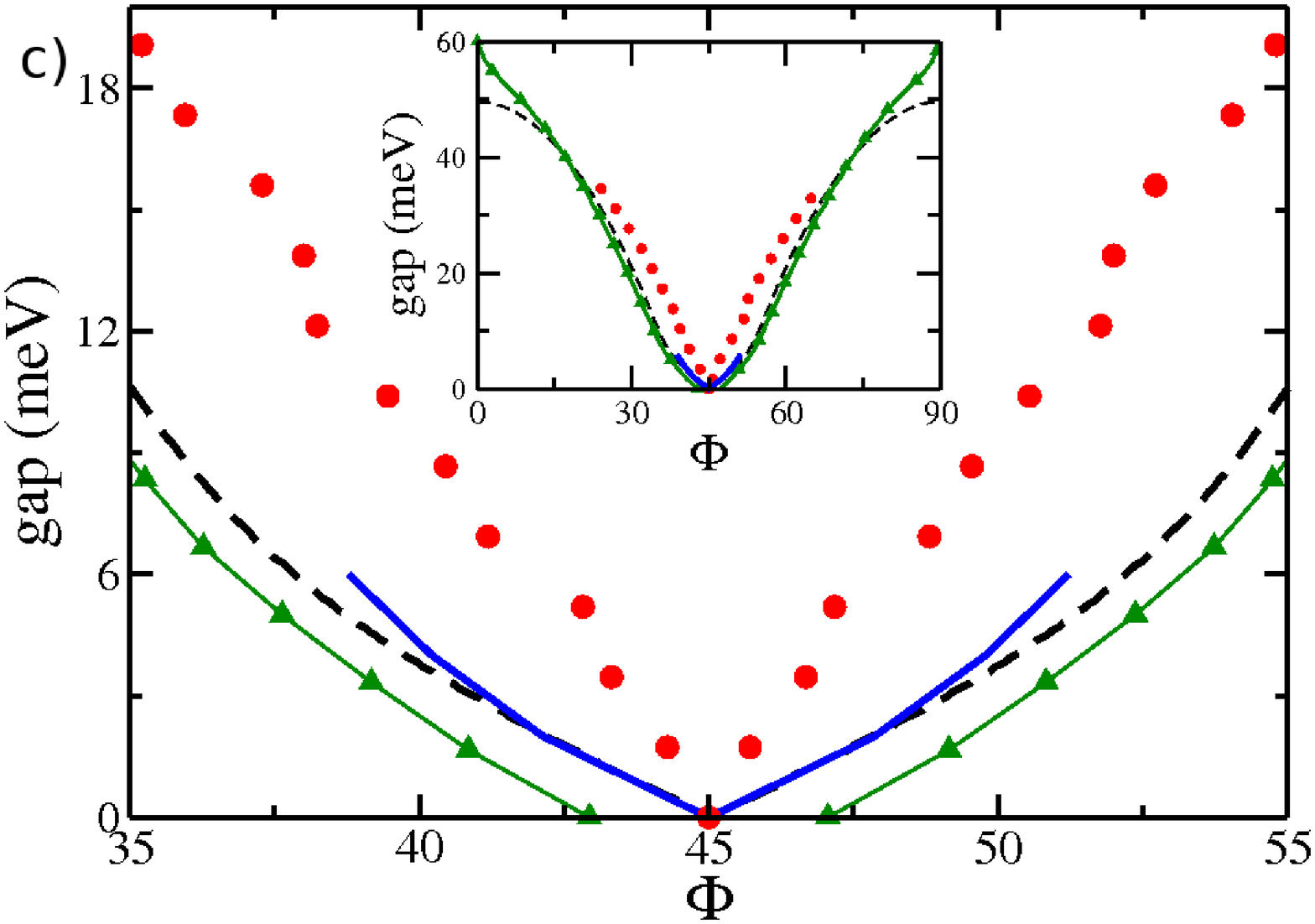} \caption {(Color online) Analysis of the moderately underdoped case.
The excitation gap $\Delta$ and order parameter $\Delta_{sc}$ are
equivalent at $T=0$. (a) The differential conductance at
$T=0.1,0.5,$ and $0.9T_c$. The arrow  identifies a slight ``kink" or
inflection point in the conductance. The inset shows the temperature
dependencies of $\Delta$ and $\Delta_{sc}$, where the shaded area
indicates the size of $\Delta_{pg}$ that contributes to the full
gap. (b)The four spectroscopic gaps extracted at $T=0.1T_c$.(c)Nodal
region of the extracted gaps at  $T = 0.9 T_c$, where various
spectroscopies yield differently shaped gaps. The inset show the
full range of the gaps. } \label{fig:modpanel}
\end{figure*}
\begin{figure*}
\includegraphics[width=2.1in,clip]
{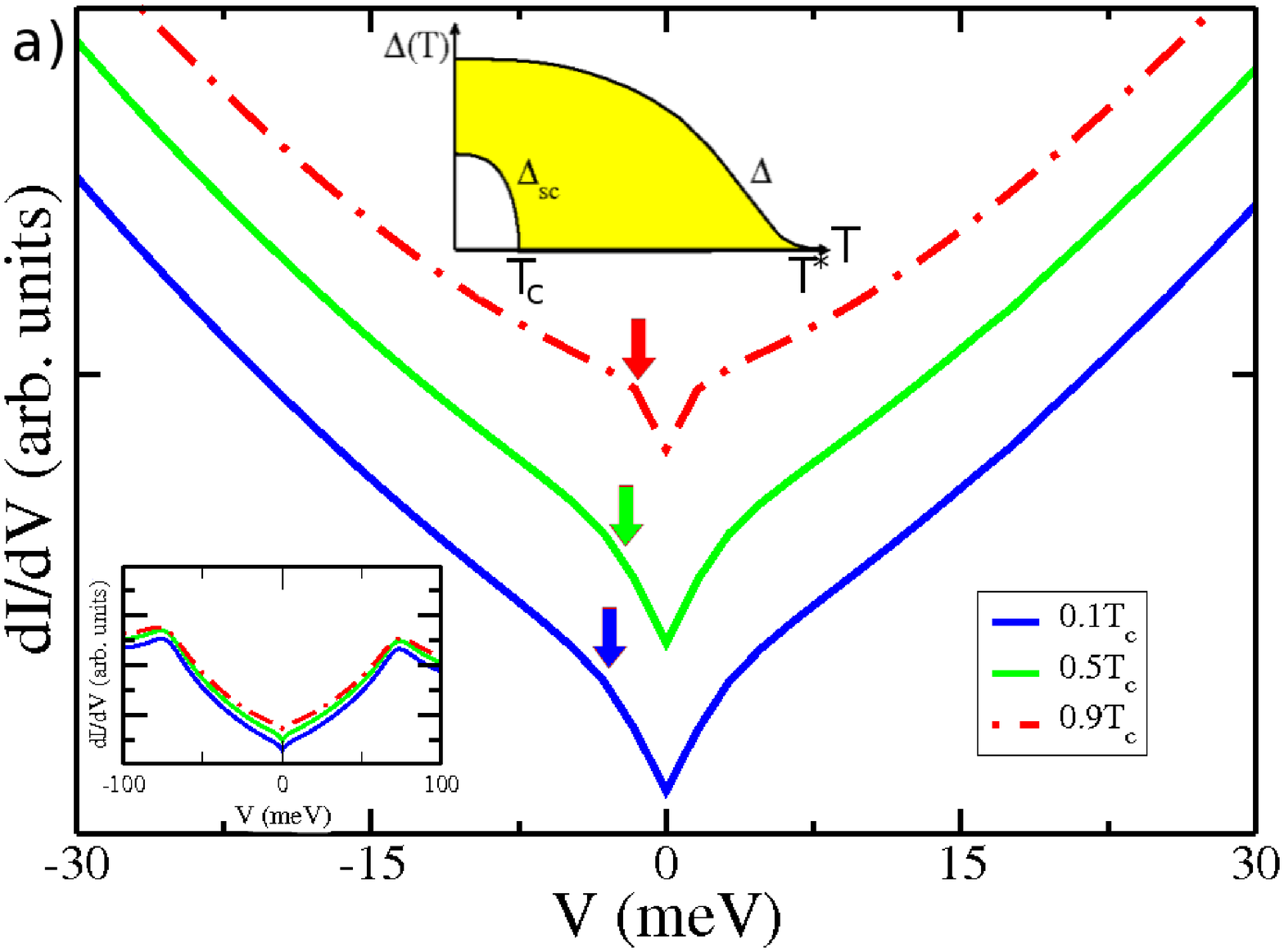}
\includegraphics[width=2.1in,clip]
{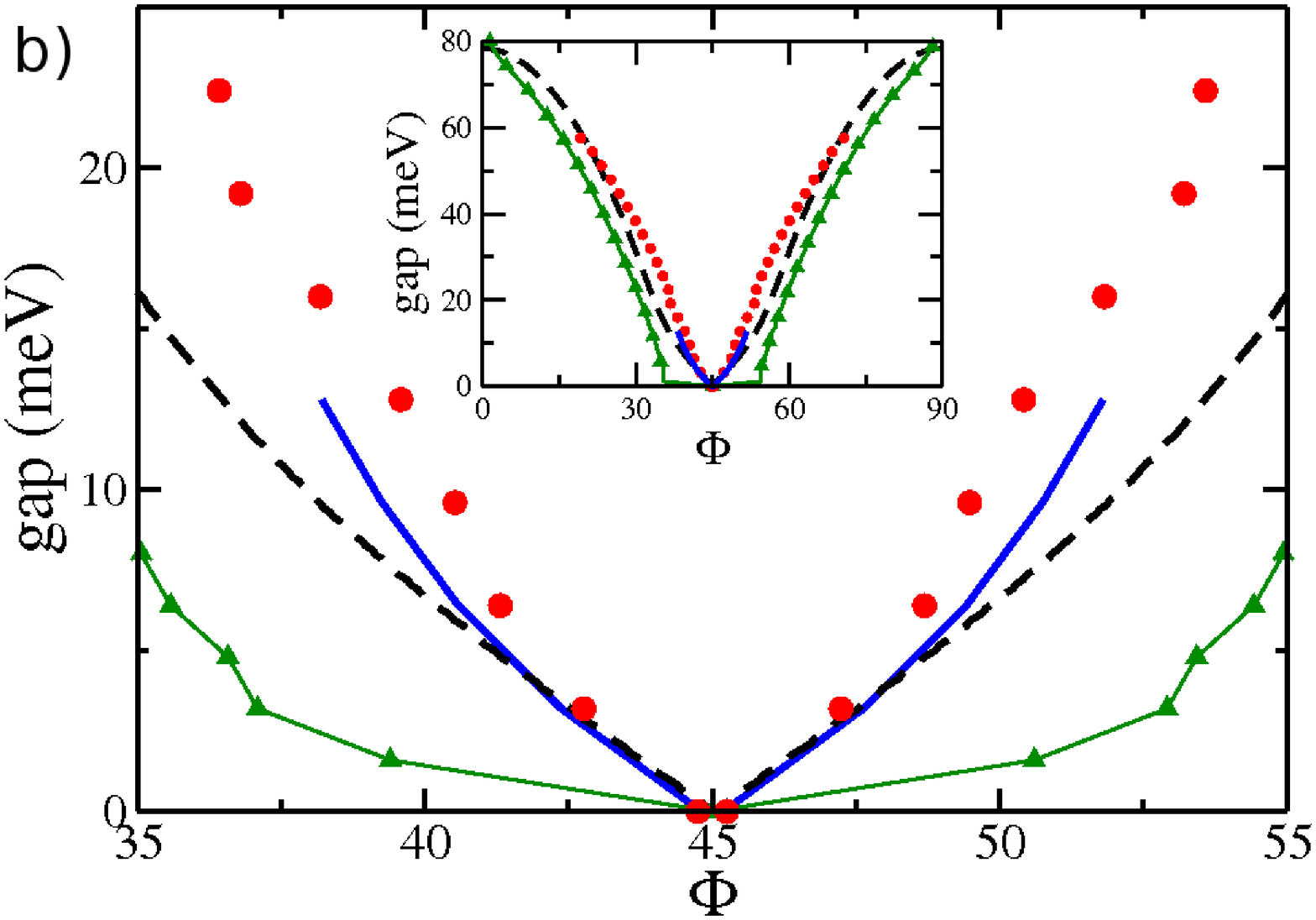}
\includegraphics[width=2.1in,clip]
{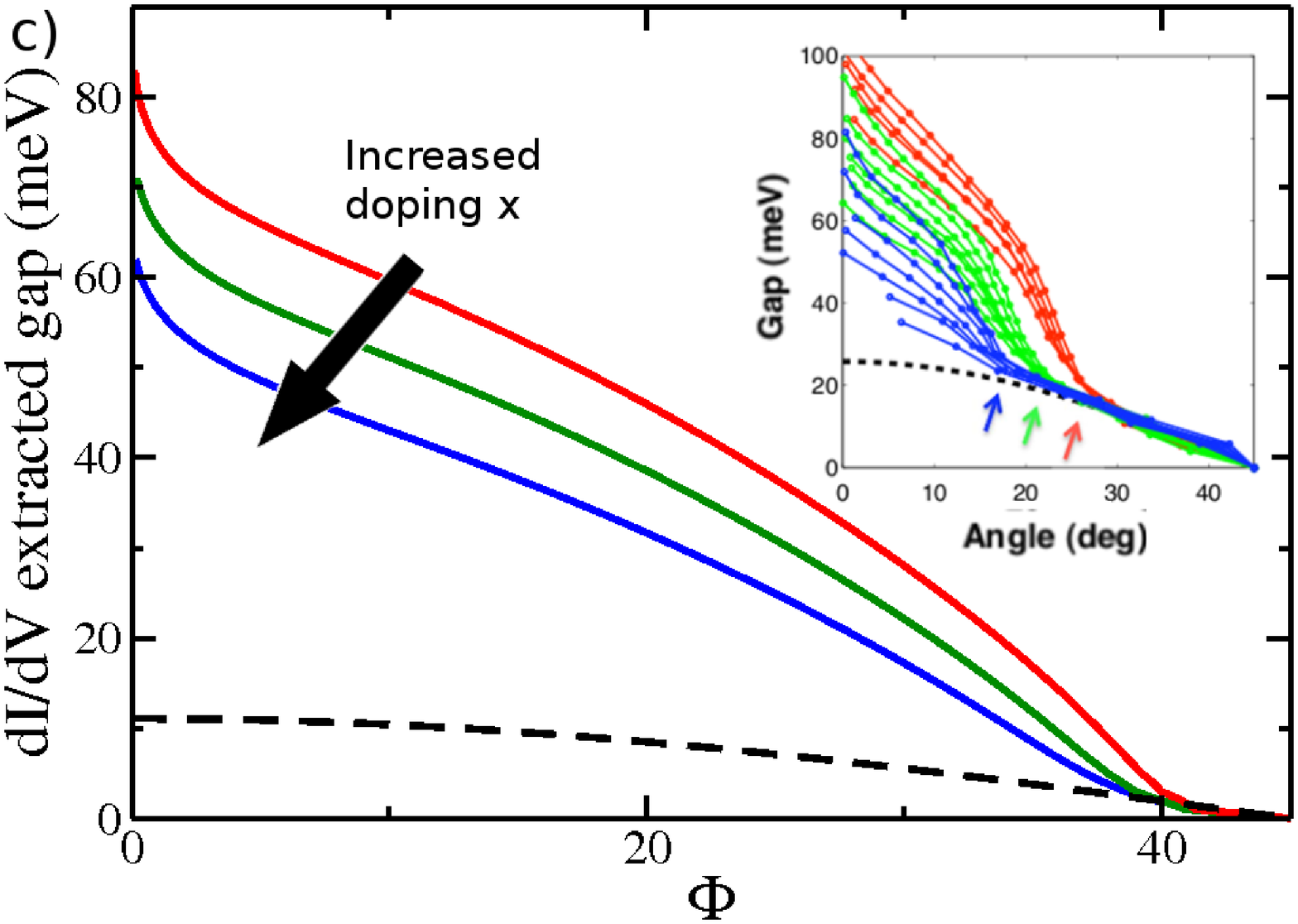}
\caption
{(Color online) Analysis of the heavily underdoped case. (a)The differential conductance at temperatures $T=0.1,0.5,$ and $0.9T_c$. The arrows indicate kinks in the gap. The lower left inset is the full range of the differential conductances. The centered inset shows the temperature dependences of $\Delta$ and $\Delta_{sc}$ in the heavily underdoped model. (b)Extracted gaps in the heavily underdoped model at $T=0.1T_c$. Different spectral gaps are labeled identically as those in Fig. \ref{fig:modpanel} (b). (c) Differential conductances at three different dopings, with
inset showing the counterpart data \cite{PuPa}. The dashed curve is a d-wave fit to the nodal regime. }
\label{fig:heavypanel}
\end{figure*}
It is important to note that $J({\bf q},\omega)$ and
$\delta n({\bf q},\omega)$ are two distinct physical quantities and
thus not directly related, since the former is a convolution of
Im$G$ only, while the latter contains both the real and imaginary
parts of $G$ and $F$.

The deduction of the associated spectral gaps proceeds as follows.
In ARPES experiments, we deduce a spectral gap from the peak to peak
separation of the symmetrized spectral function \cite{arpesRMP}. The approach of
extracting a gap from the differential conductance is
outlined in Ref.~\onlinecite{PuPa}, where
the dI/dV curve is fitted with a sum of BCS-like
gaps $\Delta_j$ with corresponding weights $W_j$, according to
\begin{equation}
\frac{dI}{dV}(V)\propto\displaystyle{\int}d \omega f^{\prime}(\omega+V)\displaystyle{\sum_{j}}\textrm{Re}\frac{\omega-i\Gamma}
{\sqrt{(\omega-i\Gamma)^2-\Delta_j^2}}W_j
\end{equation}
The parameter $\Gamma$ is used to fit the differential conductance
at zero bias.  Once the weights are obtained, they implicitly define
the angular dependence of the gap, $\Delta(\Phi)$, from the inversion
of $\Phi(\Delta=\Delta_j)\propto\sum_{i=1}^jW_i$.

The extraction of the spectral gaps both from $J({\bf q},\omega)$
and $\delta n({\bf q},\omega)$ assumes the validity of the octet
model \cite{Seamus}. This model is based on the observation that, since the
spectral function $A(\textbf{k},\omega)$ is peaked at the tips of
the contours of constant energy $\omega$, the convolution of two
spectral functions in $J({\bf q},\omega)$ should exhibit peaks at
those momenta $\textbf{q}_i$ that connect the tips. The resulting
dependence of $\textbf{q}_i$ on energy,
$\textbf{q}_i(\omega=\Delta(\Phi))$, can then be inverted to obtain the
momentum (or angular) dependence of the gap. While $\delta n({\bf
q},\omega)$ is not a convolution of two spectral functions, it has
been suggested that a similar argument still holds. 
We anticipate that 
since the octet model is derived from considerations of $J(\textbf{q},\omega)$,  its related inversion procedure will yield
a spectral gap closer to a pure $d$-wave form.
In our 
extraction of the spectral gaps, we use the octet vectors
$\textbf{q}_1$ and $\textbf{q}_5$, and $\textbf{q}_3$ and
$\textbf{q}_7$. Each pair of vectors yields a different angle
function $\Phi(\Delta)$ and we require for consistency of the octet
model that the two angular functions differ by no more than
$5\%$.

We start by discussing the spectral gaps in the moderately
underdoped cuprates. In the first (second) column of
Fig.\ref{fig:bigpanel}, we present contour plots of $\delta n({\bf
q},\omega)$ ($J({\bf q},\omega)$) at $\omega=-10meV$. From top to
bottom, the rows correspond to temperatures $T=0.1, 0.9,$ and
$1.1T_c$. For $T=0.1T_c$ all octet vectors can be
identified in the contour plots. As the temperature is increased and non-condensed pairs
(with non-vanishing $\gamma$) replace the coherent superconducting
pairs, the intensity of the peaks rapidly reduces in both
spectroscopies. The intensity of the octet vectors eventually
vanishes above $T_c$, signaling the breakdown of the superconducting
state. There is a key distinction between the $\delta n({\bf
q},\omega)$ and $J({\bf q},\omega)$. The analysis of $\delta n({\bf
q},\omega)$ \cite{ourQPI} shows an extinction point (the energy
above which the octet vectors are no longer consistent) that scales
with the order parameter, $\Delta_{sc}$, and thus terminates at
$T_c$. It should be noted that this result appears to differ with
some recent experiments on heavily underdoped cuprates \cite{Seamus,
DavisScience09} associating the QPI extinction with the
antiferromagnetic zone boundary and finding an, albeit greatly
reduced, QPI signal above $T_c$. This latter, controversial
\cite{Priv}, point does not appear \cite{QPIFail} to apply to the
moderately underdoped cuprates. 
In contrast, the octet vectors in $J({\bf
q},\omega)$ are more robust and persist to energies comparable to
the total gap ($\approx 0.8\Delta$). A small trace of the $J({\bf
q},\omega)$ octet vectors is found above $T_c$, but the peaks are
substantially broadened and the octet inversion is problematic, as
is consistent with some experimental claims \cite{JDOSPSEUDOGAP}.

In Fig. \ref{fig:modpanel}(a) we plot the differential conductance
at three different temperatures ($T=0.1,0.5,$ and $0.9T_c$) in the
moderately underdoped system. At the lowest temperature $T=0.1T_c$,
the behavior is characteristic of a simple $d_{x^2-y^2}$-wave gap.
As the number of non-condensed pairs increase with increasing
temperature, the coherence peaks broaden. The
red arrow denotes a slight kink in the shape of the $dI/dV$ spectrum
which appears near $T_c$.
The inset plots the temperature dependence of the superconducting
gap, $\Delta_{sc}$, which vanishes at $T_c$, and total gap $\Delta$,
which persists up to $T^*$. A collection of the extracted gaps for $T =
0.1 T_c$ in Fig.\ref{fig:modpanel}(b) shows that they are in good
quantitative agreement over a wide range of angle $\Phi$. However,
the gaps obtained from $\delta n({\bf q},\omega)$ and $J({\bf
q},\omega)$ exhibit different extinction energies near $35$ and $45
meV$, respectively. In contrast, for $T=0.9T_c$, the agreement
between the various gaps breaks down, primarily in the nodal region,
as shown in Fig.\ref{fig:modpanel}(c).  Except for
$J({\bf q},\omega)$, all gaps show a suppression in the nodal region
associated with the diminishing $\Delta_{sc}$, with the largest
suppression found in the gap extracted from the $dI/dV$ fit. 
This behavior is associated with the low energy part of the $dI/dV$ curve,
where the slope, for
$V\approx -3$ to $3 meV$ is larger than elsewhere, 
resulting in a smaller gap.

We note that the
discrepancies between these four spectral gaps scale with the
deviation of the gap from a pure $d_{x^2-y^2}$-form. 
These gap deviations at high temperatures below $T_c$ are not
unexpected in view of the second order phase transition and the
presence of Fermi arcs above $T_c$\cite{KaCh}. Indeed, this smooth evolution
with decreasing $T$ of the ARPES spectral gap from a Fermi arc shape
first to a distorted $d$-wave (near $T_c$) and then to a simple
$d$-wave ground state has been reported \cite{KaCh,ShenNature}
for moderately underdoped cuprates. At a
microscopic level, these deviations from a simple $d$-wave are
associated with the relatively small size of $\Delta_{sc}/\Delta$.

In developing a model for the strongly underdoped cuprates, we were
guided two experimental observations at $T \ll
T_c$: (i) the $dI/dV$ and ARPES - derived nodal gaps show a strong
suppression \cite{PuPa,ShenNature}, and (ii) the differential conductances exhibit
inflection points (kinks)\cite{PuPa}. The results suggest that
both observations can be explained within our theoretical scenario
if one assumes that even in the limit $T \rightarrow 0$, an
appreciable fraction of non-condensed pairs is present, perhaps due to
a contamination from the insulating phase. A schematic temperature
dependence of the total gap, $\Delta$ and the order parameter
$\Delta_{sc}$ is shown in the lower left inset of Fig.
\ref{fig:heavypanel} (a). For concreteness, we take
$\Delta_{sc}(0)/\Delta_{pg}(0)=0.5$; the details of the temperature
evolution of the gaps are outlined in Ref. \onlinecite{ourarpes}.
The resulting differential conductance for three different
temperatures ($T=0.1,0.5$, and $0.9T_c$) is shown in Fig.
\ref{fig:heavypanel}(a). The lower right inset plots the frequency
dependence of $dI/dV$ over a wider range.

In Fig. \ref{fig:heavypanel} (b), we present the spectral gaps
extracted from the four spectroscopic techniques for $T=0.1T_c$. We
again observe that the deviation of the gap from a pure
$d_{x^2-y^2}$-form coincides with significant discrepancies between
these four spectral gaps. The nodal region, shown in the main figure
emphasizes \textit{the clear similarity between the gaps extracted
from the low temperature heavily underdoped model (Fig.\ref{fig:heavypanel}(b))
and those near
$T_c$ in the moderately underdoped samples (Fig.\ref{fig:modpanel}(c))}.
The gaps over the full
range of angles are shown in the inset. Finally, in Fig.
\ref{fig:heavypanel} (c) we present the evolution of $dI/dV$ with
doping, which was experimentally investigated in
Ref.\onlinecite{PuPa} (see inset). Here, we hold the $T=0$
superconducting gap $\Delta_{sc}$ constant while the zero
temperature pseudogap $\Delta_{pg}$ is taken to increase as the
doping is decreased, thereby leading to an increased total gap
$\Delta$ with underdoping, as experimentally observed. In accord
with experiment \cite{PuPa}, there is universal behavior near $V=0$
associated with the superconducting contribution while the spectra
for the lowest (highest) doping breaks away from the universal curve
at the smallest (highest) $\Phi$.

Past work in the
literature \cite{ViNowa,DavisNature2003,JDOSPSEUDOGAP} generally
builds on the premise that there is universality in the spectral gap
measurements. In Ref.~\onlinecite{McEGwe} good agreement was reported between $\delta
n({\bf q},\omega)$ and $J({\bf q},\omega)$ (albeit in a case where
there are no deviations from the pure $d$-wave gap shape). In Ref.~\onlinecite{ViNowa} it was presumed that coherence established in ARPES
experiments near the antinodes would reflect on the extinction point
found in QPI, and that such extinction was difficult to understand
without invoking specific impurity models.  By contrast, in this paper we have addressed four
different spectroscopic probes in one theoretical framework and
shown how disparate the results are, except in the limiting case of
low $T$ moderately underdoped cuprates.
Another important conclusion of the present
approach is that it is sufficient to presume a
smooth evolution of the spectral gaps with
doping $x$ and temperature $T$ in order to understand
the heavily underdoped cuprates; introduction of a quantum critical point or competing order is not necessary.

This work was supported by Grant Nos. NSF PHY-0555325 and NSF-MRSEC
DMR-0213745 and by the U.S. Department of Energy under Award No.
DE-FG02-05ER46225 (D.M.). We thank A. Yazdani and C. Parker
and M. Hashimoto for
their help and advice.

\bibliographystyle{apsrev} 
\bibliography{Review2}

\end{document}